\documentclass[pra,floatfix,twocolumn,showpacs]{revtex4-1}
\usepackage{amsmath}
\usepackage{graphicx}
\usepackage[usenames,dvipsnames]{color}
\usepackage[colorlinks,linkcolor=blue,citecolor=blue,urlcolor=blue]{hyperref}

\begin{document}
\title{Localization due to interaction-enhanced disorder in bosonic systems}
\author{Rajeev Singh and Efrat Shimshoni}
\address{Department of Physics, Jack and Pearl Resnick Institute, Bar-Ilan University, Ramat-Gan 52900, Israel.}
\date{\today}
\begin{abstract}
Localization in interacting systems caused by disorder, known as many-body localization (MBL), has attracted a lot of attention in recent years.
%
%
Most systems studied in this context also show single-particle localization, and the question of MBL is whether the phenomena survives the effects of interactions.
It is intriguing to consider a system with no single-particle localization but which does localize in the presence of many particles.
The localization phenomena occurs ``due to'' rather than ``in spite of'' interactions in such systems.
%
We consider a simple bosonic system and show that interactions enhance the effects of very weak disorder and result in localization when many particles are present.
%
We provide physical insights into the mechanism involved and support our results with analytical and numerical calculations.
\end{abstract}
\maketitle
\section{Introduction}
The effect of disorder on transport properties of a quantum system is an old problem which has reinvigorated a lot of interest lately under the name of many-body localization (MBL)~\cite{Anderson1958,Evers2008,Basko2006,Oganesyan2007,Pal2010,Bardarson2012,Luitz2014,Singh2015,Ovadia2014,Schreiber2015,Smith2015,Bordia2016,Choi2016,Bordia2016a,Nandkishore2014,Altman2014,Vasseur2016}.
This recent effort is focused on understanding the effect of disorder in the presence of interactions.
A curious idea which has emerged is whether disorder is at all needed for the localization phenomena~\cite{Schiulaz2013,Grover2013,Hickey2014,DeRoeck2014,DeRoeck2014a}.
One would expect that localization in the absence of disorder is not possible due to many-body resonances, and any observed localization-like effect in dynamics must be transient~\cite{Yao2014b}.
On the other hand bosonic systems, in contrast to fermionic ones, have the possibility of many particles occupying the same state, which can lead to dramatic effects of interactions.
The recent work of \textcite{Pino2015} considers a strongly interacting bosonic system, and shows semi-classically as well as in a full quantum description that this system exhibits a departure from conventional statistical physics behavior.
This breakdown of thermodynamic description has been called MBL even though there is no disorder.
Studies of localization in absence of disorder have given rise to another interesting direction: systems with extended single-particle states but exhibiting localization for many particles~\cite{Sierant2016,BarLev2016}.
In these studies the disorder is introduced in the interaction strength and hence has no effect on single-particle properties.
These systems show localization because of interactions and are appropriately termed MBL.
However these systems do have high disorder at the many-particle level.
%
%

%
In this work we take yet another direction and consider the response of a bosonic system to a seemingly very weak disorder, such that the single-particle localization length is much larger than the system under consideration.
Hence in the absence of interactions, the system is effectively delocalized.
We then study how the presence of interactions among the particles enhances this weak disorder and causes localization in the many-body states.
This effect is the same as macroscopic self-trapping~\cite{Smerzi1997,Albiez2005}, where an isolated system in a high energy state can not relax due to the absence of channels that can take away the excess energy.
To this end, we consider one of the simplest systems, namely the two-site Bose-Hubbard (BH) model~\cite{Sachdev2011, Morsch2006, Bloch2008, Milburn1997, Castin1997, Sakmann2009, Boukobza2009, Chuchem2010, Simon2012, Veksler2015}, that captures the essence of localization phenomena by interaction-enhanced disorder while being amenable to some analytical insights into the physical mechanism.
We analyze the eigenstates and energy levels of the system and also calculate various physical quantities numerically to show the localization-delocalization phase diagram.
%
We provide physical arguments in terms of resonances between energy levels and arrive at the condition for localization analytically.
Our numerical calculations are then extended to slightly larger number of sites, yielding overall features that are remarkably similar to the two-site case.
Although the large eigenstate to eigenstate fluctuations for the bigger system do not allow us to draw the corresponding phase diagram, qualitative effects of our simple arguments can be clearly seen.
We present a discussion about how the physical insights obtained for this simple system can be generalized, and lie at the heart of the localization phenomena due to interaction-enhanced disorder even for larger systems.
The paper is organized as follows.
In section~\ref{model} we introduce the model and discuss some of its properties and the numerical results showing the localization-delocalization phase diagram for the system with two sites.
We also present the physical argument of resonance of energy levels for the two-site BH model and derive the analytical expression for the phase diagram in this section.
We then show the numerical results for slightly larger systems.
In section~\ref{LargerSystem} we outline the physical picture about how this phenomena might be manifested for longer chains.
We summarize the results and provide an outlook in section~\ref{Conclusions}.
An analytical derivation of the onset of quantum effects is presented in Appendix~\ref{StrongInteraction}.

\section{Model and Main Results} \label{model}
We consider a system of $N$ bosons described by the Bose-Hubbard (BH) model on a 1d chain~\cite{Sachdev2011, Morsch2006, Bloch2008}
\begin{equation} \label{Hamiltonian}
H = \sum_i \left[-J (b_i^\dagger b_{i+1} + h.c.) + \frac{U}{2} n_i (n_i-1) + V_i n_i \right]
\end{equation}
where $J$ is the hopping amplitude (kinetic energy coefficient), $U$ is the on-site interaction between the bosons, $V_i$ is the local potential on each site and $n_i \equiv b_i^\dagger b_i$ is the number operator on site $i$.
This model has been studied in a wide range of contexts ranging from superfluidity to quantum chaos~\cite{Sachdev2011, Morsch2006, Bloch2008}.
Its low-energy properties are well-known: the ground state of the system exhibits a superfluid to Mott insulator transition with increasing interaction strength.
The emphasis of MBL on the other hand is on the {\em high-energy} states where a similar transition might occur.
For high energies, the delocalized and localized states can be thought of as the generalization of the superfluid and Mott phases respectively.
As detailed below, we perform an analytical treatment of this system with just two sites, i.e. the two-site BH model, and present numerical results for two-, three- and four-site versions of the system.
We will also present a qualitative discussion of the more general case of an arbitrary lattice.
Since we are interested in localization-delocalization transition in this system, which can be thought of as the study of the relative importance of quantum effects, we will perform a full quantum treatment of the problem even when the semi-classical approximation is justified.
%
%
In a very simplified setting we will show how this system is capable of showing qualitatively different behavior in different eigenstates depending on the energy.
This feature lies at the heart of the localization phenomena, and the system indeed has a mobility edge even with very weak disorder when large number of particles are present.

\subsection{Two-site Bose-Hubbard model: numerical results}
\begin{figure}
\includegraphics[width=0.99\linewidth]{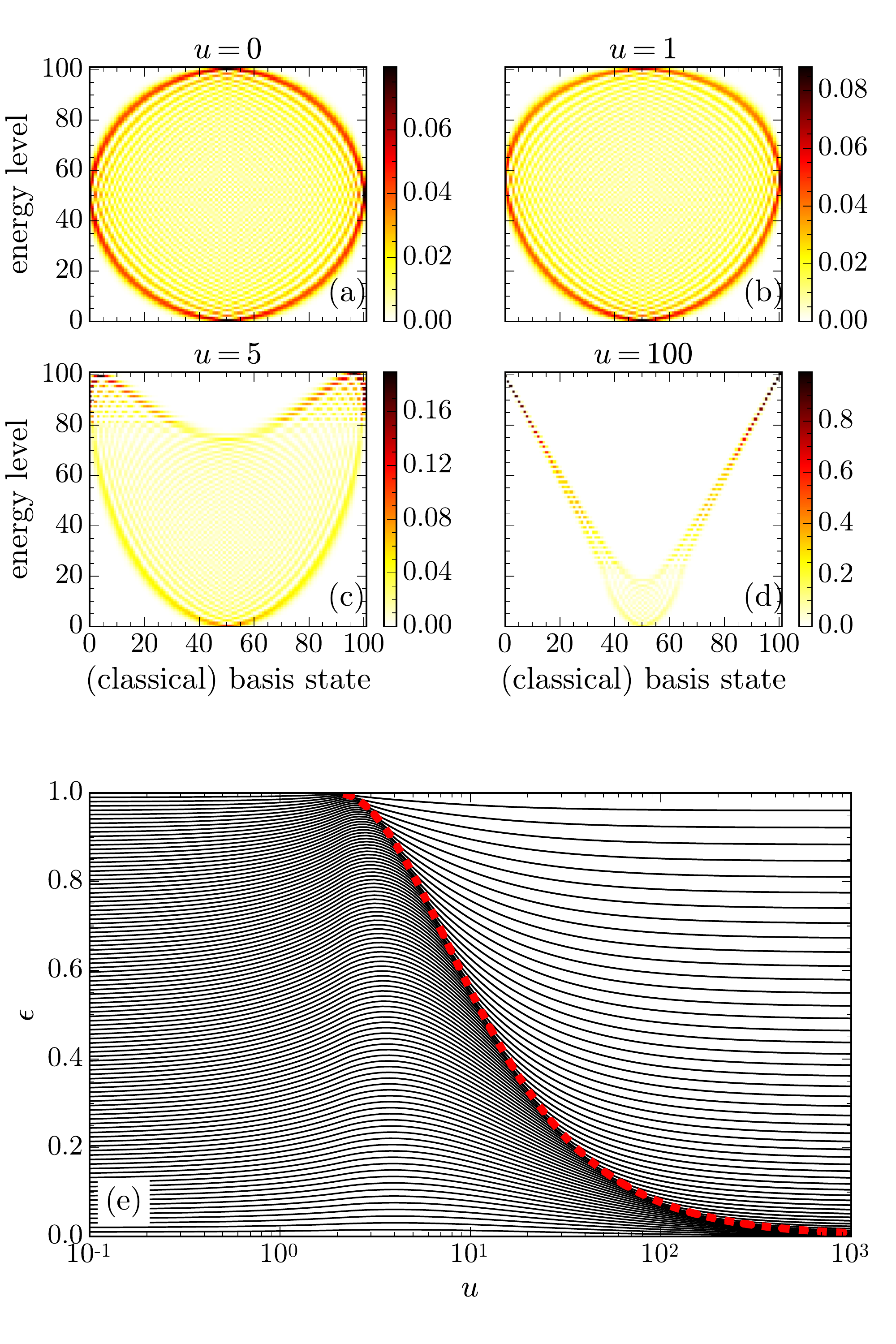}
\caption{Eigenstates for the two-site Bose-Hubbard system with $100$ bosons at different values of scaled interaction strength $u = NU/J$ (a-d) and a small symmetry-breaking term $V=10^{-6}$.
The basis is chosen to be the ``classical'' one where the states are labeled by the number of bosons on one of the sites.
(e) Normalized energy $\epsilon$ as a function of $u$ captures the localization-delocalization phase diagram.
The red dashed curve corresponds to the equation of separatrix Eqn.~\ref{SeparatrixEquation} for the model in the semi-classical analysis.
}
\label{TwoSiteEigenstates}
\end{figure}
\begin{figure*}
\includegraphics[width=0.99\linewidth]{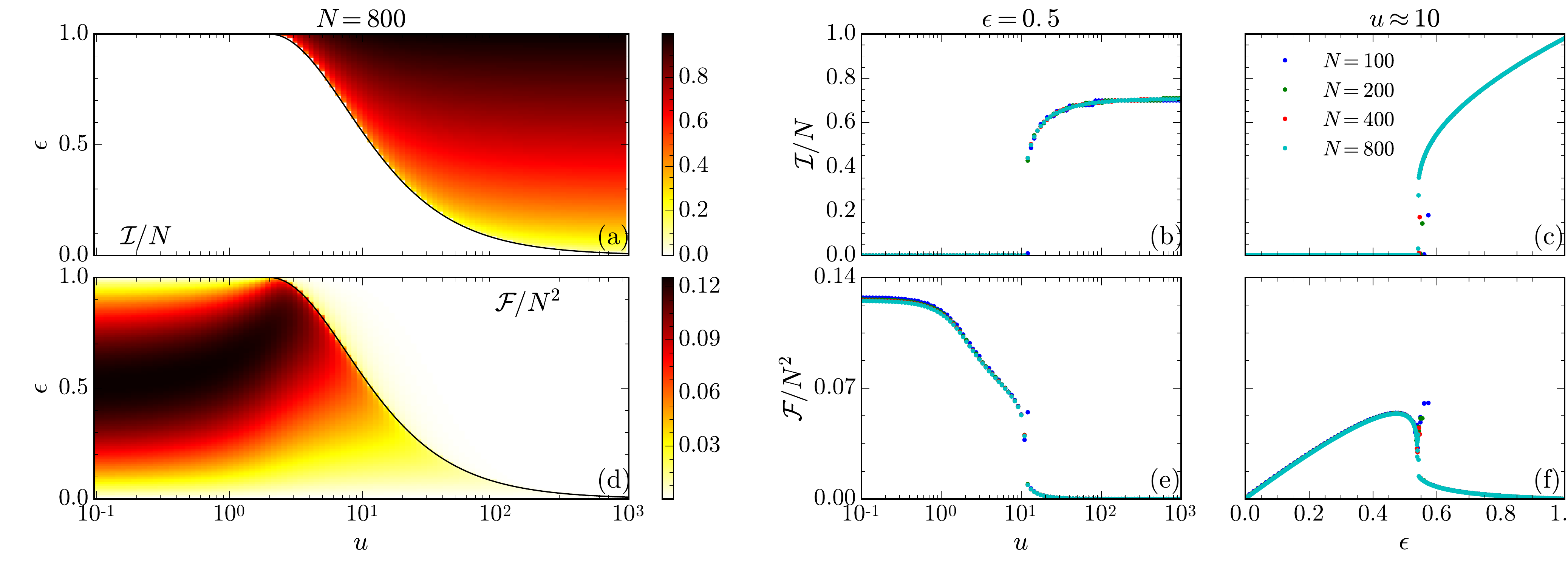}
\caption{Imbalance $\mathcal{I}$ and particle number fluctuations $\mathcal{F}$ with normalized energy $\epsilon$ and the relative interaction strength $u=NU/J$ for the two-site Bose-Hubbard system.
(a) shows the imbalance for $N=800$ using a false color plot as a function of $\epsilon$ and $u$.
(b) and (c) show vertical and horizontal cuts in (a) for different number of particles ($N=100,200,400,800$) as a function of $u$ (at $\epsilon=0.5$) and $\epsilon$ (at $u \approx 10$) respectively.
Similarly (d) shows the particle number fluctuations for $N=800$ as a function of $\epsilon$ and $u$, and (e) and (f) show the the cuts in (d) as a function of $u$ (at $\epsilon=0.5$) and $\epsilon$ (at $u \approx 10$) respectively.
The black solid line in (a) and (d) is the semi-classical equation of the separatrix Eqn.~\ref{SeparatrixEquation} for the system.
}
\label{TwoSiteObservables}
\end{figure*}

The two-site Bose-Hubbard (BH) model~\cite{Milburn1997, Castin1997, Sakmann2009, Boukobza2009, Chuchem2010, Simon2012, Veksler2015}
\begin{equation} \label{TwoSiteBoseHubbard}
  H = -J \left(b_1^\dagger b_2 + h.c. \right) + \sum_{i=1,2} \frac{U}{2} n_i (n_i-1) + V n_2
\end{equation}
is integrable.
In the semi-classical limit~\cite{Chuchem2010}, the phase space is two-dimensional and since the system follows a Hamiltonian dynamics the energy determines the trajectory, which is reminiscent of the dynamics of a pendulum.
This analysis suggests different regimes of the system with respect to the model parameters, which can be combined to form a single dimensionless one
\begin{equation}
u = \frac{NU}{J}.
\end{equation}
When $u<2$ the system is in the so-called Rabi regime, while for $2< u \ll 2N^2$ it is in the Josephson regime and for $u>2N^2$ in the Fock regime~\cite{Chuchem2010}.
%
%
In the localization language, the three regimes correspond to fully delocalized, partially localized with a mobility edge, and fully localized phases respectively.
In Fig.~\ref{TwoSiteEigenstates}~(a-d) we show the effect of interaction on eigenstates of the two-site BH system with $N=100$ particles for $J=1$ and $V=10^{-6}$.
Note that we assume a very small energy difference $V$ between the two sites (mimicking a weak disorder potential), whose significance will become clear later.
We use the ``classical'' states (characterized by a well-defined number of bosons on each site) given by
\begin{equation} \label{ClassicalStates}
|m\rangle = \frac{(b_1^\dagger)^m (b_2^\dagger)^{N-m}}{\sqrt{m! (N-m)!}} |0\rangle
\end{equation}
as the basis, and calculate the overlap with each eigenstate.
The classical states are the eigenstates in the limit $u \rightarrow \infty$.
In contrast, for the non-interacting case ($u=0$) all eigenstates are approximately symmetric around $m=N/2$ (Fig.~\ref{TwoSiteEigenstates}~(a)).
As $u$ is increased, the states with the highest energy (near the top of each panel in Fig.~\ref{TwoSiteEigenstates}~(a-d)) begin to broaden in this basis while still maintaining the symmetry around $m=N/2$ upto $u \approx 2$ (Fig.~\ref{TwoSiteEigenstates}~(b)).
For $u \gtrsim 2$ the states with the highest energy break this symmetry in pairs, with either state in the pair being approximately the reflection of the other around $m=N/2$ (Fig.~\ref{TwoSiteEigenstates}~(c)).
For $u \approx N$ most of the high energy states have gone through this symmetry breaking (Fig.~\ref{TwoSiteEigenstates}~(d)).
This symmetry breaking in the eigenstates is the macroscopic self-trapping or localization phenomena.
This effect is also clearly reflected by the energy spectrum itself in this simple system in terms of the (approximate) degeneracy structure.
In Fig.~\ref{TwoSiteEigenstates}~(e) we show the normalized energies
\begin{equation} \label{NormalizedEnergy}
\epsilon_i \equiv \frac{E_i - E_0}{E_\infty - E_0}
\end{equation}
as a function of $u$, where $E_0$, $E_i$ and $E_\infty$ are ground state, $i$-th state and top state energies respectively.
We clearly see a qualitatively different behavior in different parts of the phase diagram.
In terms of energy levels, the two limiting cases of the Hamiltonian Eqn.~\ref{TwoSiteBoseHubbard} are very different: in the limit $U\rightarrow 0$, which we call quantum, all states are non-degenerate whereas in the opposite limit of $J\rightarrow 0$ all states, except the ground state depending on the number of bosons, are approximately two-fold degenerate.
We call the second limit classical, as the energy eigenstates are very close to the eigenstates of the number operators.
In the Josephson regime ($2 \lesssim u \lesssim 2N^2$) the eigenstates can show both types of behavior at different energies.
For the two-site system in the Josephson regime within the semi-classical treatment this {\em mobility edge} manifests itself in the form of a separatrix, which is given by~\cite{Chuchem2010}
\begin{align}
\mathrm{ground\ energy:}\quad &E_0 = -NJ   \\
\mathrm{separatrix:}\quad &E_x = +NJ   \\
\mathrm{top\ energy:}\quad &E_\infty = \frac{1}{2} \left( \frac{u}{2} + \frac{2}{u} \right) NJ.
\end{align}
In terms of the normalized energy $\epsilon$ (Eqn.~\ref{NormalizedEnergy}), the equation of separatrix becomes
\begin{equation} \label{SeparatrixEquation}
\epsilon_x = \frac{2}{1 + \frac{1}{2} \left(\frac{u}{2} + \frac{2}{u} \right)}.
\end{equation}
We plot the separatrix energy on the energy levels phase diagram in Fig.~\ref{TwoSiteEigenstates}~(e) as the red (dashed) curve and find an excellent agreement between the full quantum calculations and the semi-classical dynamical phase diagram.
The two phases separated by the mobility edge can be considered classical and quantum in the following sense: in one of the phases (the almost degenerate one) the number of bosons on each site are almost good quantum numbers and there is little quantum fluctuations in the number of particles.
This phase can be called ``classical'', and is analogous to the localized phase, as the latter also has this property in the limit of localization length being very small (less than lattice spacing).
The other phase with non-degenerate eigenstates can be considered quantum because it has the opposite properties and the quantum fluctuations in the number of bosons on either site is non-zero.
This observation, along with symmetry breaking in the wavefunctions, motivates the following two quantities which can be used to distinguish the two phases: imbalance in the number of bosons
\begin{equation}
\mathcal{I}_{ij} \equiv |\langle \psi |(n_i - n_j)| \psi \rangle|
\end{equation}
and quantum fluctuations in the number of particles
\begin{equation}
\mathcal{F}_i \equiv \langle \psi | n_i^2 | \psi \rangle - \langle \psi | n_i | \psi \rangle^2.
\end{equation}
We show these two quantities as a function of $\epsilon$ and $u$ in Fig.~\ref{TwoSiteObservables}~(a,d) and the phase diagram can be seen very clearly.
Note that the semi-classical separatrix coincides with a sharp change in both these quantities, which is most dramatic at high $\epsilon$.
To demonstrate this, we also show horizontal and vertical cuts across the phase diagram for both quantities with different number of bosons in Fig.~\ref{TwoSiteObservables}~(b,c,e,f) and find almost no dependence on number of particles when $\mathcal{I}$ and $\mathcal{F}$ are suitably scaled.
%
%

\subsection{Two-site Bose-Hubbard model: perturbation theory and level hybridization}
We have seen in the previous subsection an excellent agreement between the mobility edge in a full quantum calculation and the equation for separatrix in the semi-classical analysis of the two-site BH system.
It is desirable to obtain this result using quantum considerations beyond this approximation.
%
%
We now derive the equation for the mobility edge for the two-site BH model using perturbation theory and a level hybridization argument.
%
%
We start in the limit of no interactions ($U=0$) and consider interactions in perturbation theory.
For $U=0$, the many-particle eigenstates are given in terms of the single-particle ones and are non-degenerate:
\begin{equation}
\tilde{E}_m = (2m-N)J, \qquad
 |\tilde{E}_m \rangle = \frac{(b_+^\dagger)^m (b_-^\dagger)^{N-m}}{\sqrt{m! (N-m)!}} |0\rangle
\end{equation}
where $b_\pm \approx ({b_1} \pm {b_2}) /\sqrt{2}$.
The interaction part of the Hamiltonian in terms of these non-interacting normal modes is given by
\begin{align}
H_{\mathrm{int}} =& \frac{U}{4} \left(4 {b_+^\dag} {b_-^\dag} {b_+} {b_-} + b_+^{\dag 2} b_+^{2} + b_+^{\dag 2} b_-^{2} + b_-^{\dag 2} b_+^2 + b_-^{\dag 2} b_-^{2}\right).
\end{align}
The perturbation has a diagonal component, including which the corrected energy is given by
\begin{equation} \label{PerturbedEnergy}
\tilde{E}'_m = (2m-N)J + \frac{U}{4} \left(N^{2} + 2 N m - N - 2 m^{2}\right).
\end{equation}
This correction to the spectrum is sufficient to obtain the desired result after making one observation regarding the interaction operator.
The unperturbed energy levels are equally spaced, but the diagonal correction acts differently on different levels.
In particular as $U$ is increased, it can reshuffle the original order of the levels; however, we do not observe any reshuffling numerically (Fig.~\ref{TwoSiteEigenstates}~(e)).
We note here that the interaction term can not hybridize two neighboring states ($m$ and $m\pm 1$): it preserves the parity of number of bosons in each mode.
But numerically the hybridization {\em does} happen between states with different parity.
The main insight from this calculation is that the weak disorder in the Hamiltonian (the imbalance $V$ in Eqn.~\ref{TwoSiteBoseHubbard}) provides the off-diagonal term that causes hybridization between nearest energy levels.
The picture we obtain is that the diagonal part of the interaction operator brings the energy levels close enough so that the disorder term, which can be arbitrarily small, leads to hybridization.
As in numerical calculation we can not set the disorder to zero, we will not see a level crossing between states with different parity numerically.
%
%

%
We make two assumptions to calculate the mobility edge: first that Eqn.~\ref{PerturbedEnergy} gives a reasonable estimate of the extreme energies (top and bottom).
In other words we assume that Eqn.~\ref{PerturbedEnergy} can be used to calculate the total bandwidth even in the presence of strong interactions.
And second, that the mobility edge (the energy at which the perturbation theory breaks down) remains at the same relative distance from the ground state for different $U$, i.e. $\tilde{E}'_{\mathrm{mob}} - \tilde{E}'_0 = 2NJ$.
The second assumption can be rephrased as: the bandwidth of the non-hybridized states is equal to the total bandwidth of the unperturbed (non-interacting) system.
The maximum energy in Eqn.~\ref{PerturbedEnergy} is obtained at $m = m^\star = N \left(\frac{1}{2} + \frac{2}{u}\right)$ and the equation of mobility edge becomes
\begin{equation}
\epsilon_{\mathrm{mob}} = \frac{2}{1 + \frac{1}{2} \left(\frac{u}{4} + \frac{4}{u} \right)} 
\end{equation}
which agrees with the separatrix equation Eqn.~\ref{SeparatrixEquation} after a rescaling of $u$ by a factor of $2$.
Thus we have obtained the mobility edge equation from purely quantum considerations of energy levels, which has also given us an insight on the importance of the disorder term however small.
%
%

\subsection{Three- and four-site Bose-Hubbard models: numerical results}
\begin{figure*}
\includegraphics[width=0.99\linewidth]{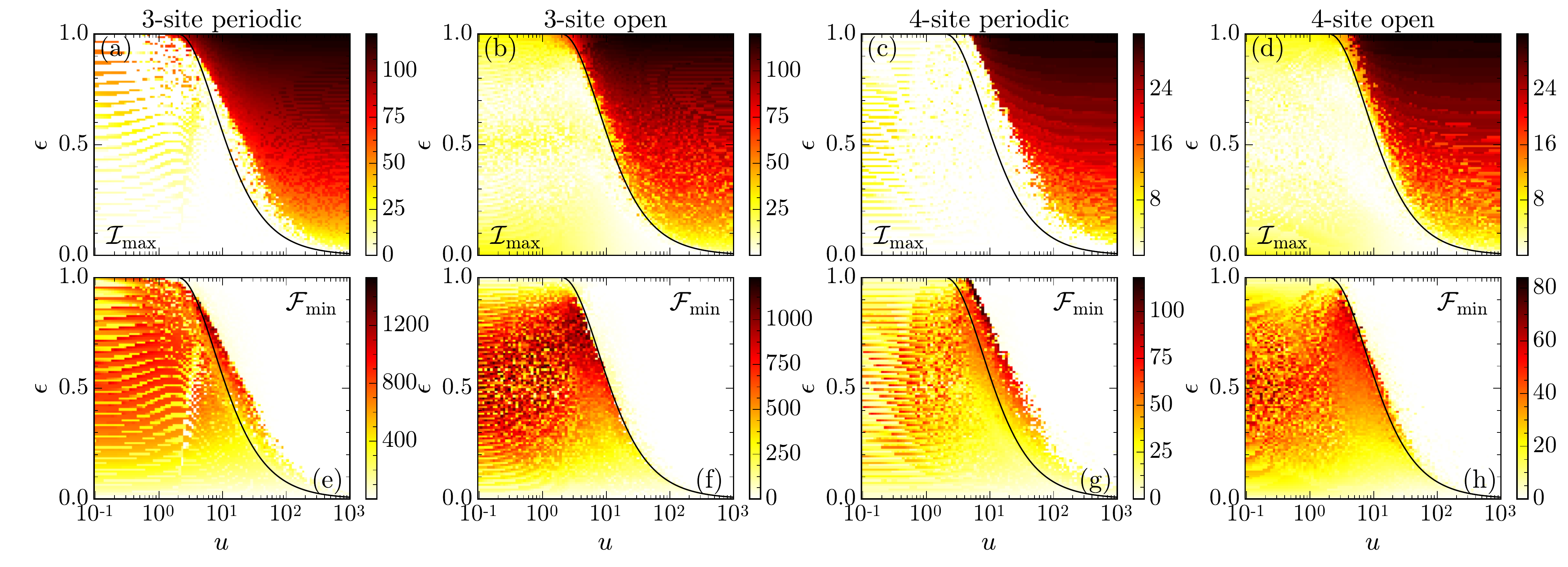}
\caption{False color plots showing maximum imbalance $\mathcal{I}_\mathrm{max}$ and minimum particle number fluctuations $\mathcal{F}_\mathrm{min}$ with normalized energy $\epsilon$ and the relative interaction strength $u=NU/J$ for the three- and four-site Bose-Hubbard systems with periodic and open boundary conditions.
Solid lines in each plot is the equation for separatrix Eqn.~\ref{SeparatrixEquation} for the two-site system from the semi-classical analysis.
}
\label{ThreeSiteObservables}
\end{figure*}
The observed phenomena described above may be special to the two-site case and stem from its integrability.
It is therefore important to consider systems larger than two sites to make sure that the observed effects are not related to integrability.
The three-\cite{Streltsov2011, Arwas2014} and four-site BH models are non-integrable and we now explore them numerically.
The Hilbert space structure of the two-site problem is quite simple in both the limiting cases, and as a result the change of energy levels with $u$ is quite instructive (Fig.~\ref{TwoSiteEigenstates}~(e)).
By contrast, for the three- and four-site problem this is no longer the case and the energy levels have many degeneracies or near-degeneracies.
Nevertheless, a qualitative change similar to the two-site case can be seen for the three- and four-sites with both periodic and open boundary conditions.
%
%
The phase diagram is quite clearly captured by the physical quantities, measured in the eigenstates, both the imbalance and particle number fluctuations.
Since we are now dealing with more than two sites, the particle number fluctuations may be different on different sites and we show the minimum among them ($\mathcal{F}_{\mathrm{min}}$).
Similarly we measure the imbalance across every bond and show the maximum ($\mathcal{I}_{\mathrm{max}}$).
The rationale for choosing the minimum fluctuation is that the transport properties are likely going to be affected the most by such sites as they would effectively block passage across them.
Similar considerations need to be taken into account for choosing maximum imbalance across bonds.
We show $\mathcal{I}_{\mathrm{max}}$ and $\mathcal{F}_{\mathrm{min}}$ for $N=120$ bosons on three sites and $N=32$ bosons on four sites for both periodic and open boundary conditions in Fig.~\ref{ThreeSiteObservables}.
Though the BH system with periodic boundary condition does not quantitatively agree with the two-site phase diagram, it shows qualitatively similar behavior (Fig.~\ref{ThreeSiteObservables}~(a,c,e,g)).
Quite remarkably, the BH model with open boundary condition agrees to a good approximation with the two-site results even quantitatively (Fig.~\ref{ThreeSiteObservables}~(b,d,f,h)).
%
%

\section{Larger systems} \label{LargerSystem}
\begin{figure}
\includegraphics[width=0.99\linewidth]{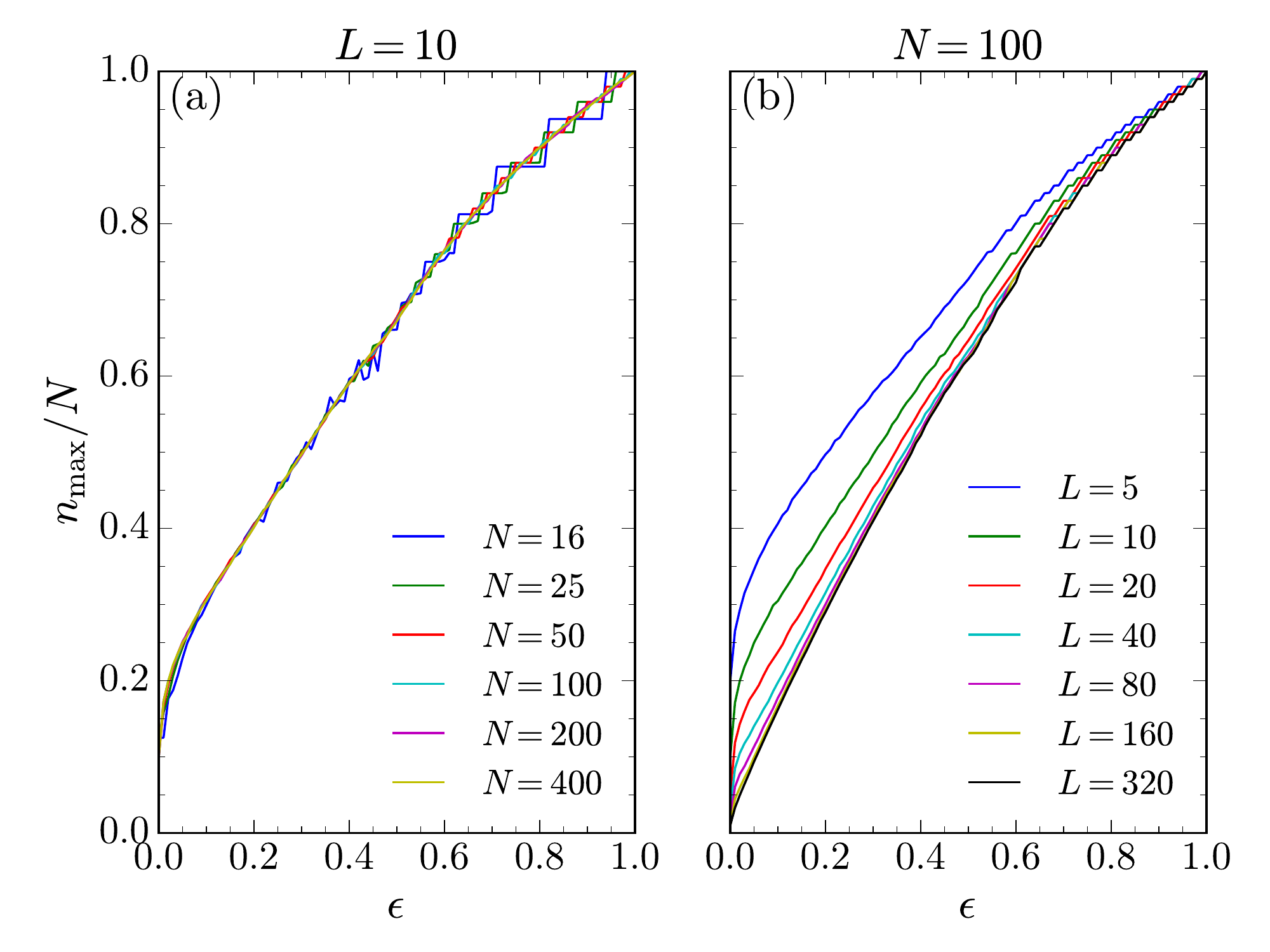}
\caption{Maximum number of bosons $n_\mathrm{max}$ on any given site along the chain as a function of normalized energy $\epsilon$ for the ``classical'' system ($J=0$).
(a) shows $n_\mathrm{max}$ normalized by total number of bosons $N$ for a given chain length $L=10$ and different $N$, while (b) shows the same quantity for a given $N=100$ and different $L$.}
\label{LongChainClassical}
\end{figure}

Due to rapid increase in the Hilbert space dimension with number of lattice sites, it is not possible to do a complete analysis of all the eigenstates for large systems numerically.
However the physical picture itself can be extended to understand their behavior and we present it here for a general lattice.
We assume a very weak disorder in the system to avoid resonances, but the disorder is assumed to be weak enough such that the single-particle localization length is larger than the system size.
There is some subtlety involved in taking the thermodynamic limit here.
%
%
If we take the limit of disorder going to zero in the thermodynamic limit such that the localization length goes to infinity as fast as the system size, it is not clear that the condition for many-body resonances over long distance will not be satisfied.
We leave this subtle issue for future works and continue with the assumption that many-body resonances can be avoided in a suitable order of limits.
The high energy eigenstates of the system must necessarily have some sites with relatively large number of bosons.
We made simple numerical estimate of the maximum number of bosons $n_\mathrm{max}$ on a site in a long chain of length $L$ (and in the limit of no hopping) as a function of normalized energy.
Interestingly, the fraction of bosons needed on one site to achieve a given normalized energy does not depend on the total number of bosons (Fig.~\ref{LongChainClassical}~(a)).
On the other hand it depends on the length of the chain quite significantly for small chains (Fig.~\ref{LongChainClassical}~(b)), but sensitivity to $L$ is reduced for $\epsilon \rightarrow 1$.
The dependence seem to approach some limiting form in the limit of longer chains.
The dynamics on sites with large number of bosons will get frozen as the kinetic term would not be sufficient to overcome the potential energy barrier needed for hopping of one boson.
The presence of a weak disorder has suppressed the many-body virtual processes that could assist such hopping via quantum tunneling.
The number of bosons per site needed for this self-trapping to occur will depend on the relative strength of the hopping and interaction parameters.
Starting from a classical state of the big system in arbitrary dimension, the two-site problem has to be analyzed for each bond in the system as a first approximation.
Every bond which has small imbalance across it will be resonant, and we can construct a network of such resonant bonds.
The state will show some finite transport if this network becomes a percolating one.
In the absence of a percolating network there will be no transport.
Regardless of whether the state has a percolating network or not, at high energy the violation of thermodynamic behavior similar to the one seen in~\cite{Pino2015} can be seen for the following reason: the sites that freeze due to self-trapping effectively remove a large number of bosons from the system.
Even if the remaining sites are well-coupled and thermalize, the system as a whole will witness poor thermalization behavior due to reduced number of bosons.
In particular the following departure from thermodynamic behavior is expected.
For a generic system, the fluctuations increase as the energy is increased.
For the system under our consideration on the other hand, an opposite behavior will be seen, as with increasing energy more bosons would enter the self-trapping phase and would not contribute to fluctuations.
These self-trapped states also exhibit the breakdown of thermalization from the point of view of eigenstate thermalization hypothesis~\cite{Deutsch1991, Srednicki1994, Rigol2008}.
A given high-energy state configuration has many permutations with very similar energies.
These permutations will have the frozen sites at different locations resulting in local observables with large deviations in eigenstates with similar energies.
%


\section{Summary and Outlook} \label{Conclusions}
To summarize, in this work we have considered a simple bosonic system and showed that interactions can enhance the effect of weak disorder and cause localization.
We have presented numerical calculations for the two-, three- and four-site systems to show this phenomena.
We rederived the semi-classical result using level hybridization arguments, which gave us a useful insight that a weak disorder is needed to understand the numerical results.
We then gave a physical picture of how this mechanism may work for larger systems.
We comment on one peculiarity of the mobility edge seen here.
The bosonic systems with repulsive interactions have a mobility edge which is very different from the usual MBL mobility edge seen in fermionic or spin systems.
Here the states {\em above} the mobility edge are localized as opposed to the ones at both ends of the spectrum for the latter~\cite{Luitz2014}.
Another difference from fermionic systems is that as this effect stems from high density of particles at a given site, it will not happen for the latter due to Pauli exclusion principle.
However for fermionic systems with many internal degrees of freedom there should be some effect of this mechanism.
Interestingly, this mechanism does not depend on whether the interaction is repulsive or attractive.
In both cases the simple physical argument goes through and we should see the localization phenomena.
However for attractive interactions the low energy states would be localized and we will obtain an inverted phase diagram.
%
%
We believe that our work will provide important physical insights to understand the more complicated systems containing arbitrary number of sites.

\section*{Acknowledgements}
%
We thank Yevgeny Bar-Lev, Richard Berkovits, Emanuele Dalla Torre, Shmuel Fishman, Aditi Mitra and Ganpathy Murthy for helpful discussions.
This work is supported by the Israel Science Foundation, grant N. 231/14.
RS also thanks the Israel Council for Higher Education's Planning and Budgeting Committee (CHE/PBC) and the Israel Science Foundation grant N. 1452/14 for financial support.

\appendix
\section{Kinetic term as a perturbation} \label{StrongInteraction}
Interestingly, performing perturbation analysis with the kinetic term is relatively simple and still captures the onset of quantum fluctuations quantitatively.
%
%
For $J=0$, the eigenstates of the system are the same as that of the particle number operators on each site which we have called classical states (Eqn.~\ref{ClassicalStates}).
%
The classical state $|m\rangle$ has energy $E_m = [m(m-1)+(N-m)(N-m-1)] U/2$.
Every energy level is almost two-fold degenerate, except the ground state which has this degeneracy only if $N$ is odd.
This almost degeneracy for the high energy states is not very relevant for the rest of the analysis as the degenerate states are not connected by the kinetic term to small order in perturbation theory.
Let us consider the situation when $m \le N/2$ with $m=0$ being the most excited state and $m=N/2$ ($(N-1)/2$) being the ground state for even (odd) $N$.
The energy spectrum is not equally spaced, which results in an interesting effect that not all eigenstates hybridize together.
The level spacing given by
\begin{equation}
\delta E_m \equiv E_m - E_{m+1} = (N-2m-1) U/2,
\end{equation}
increases linearly with energy level ($m=0$ is the highest energy level).
This spacing has to be compared to the energy due to the kinetic term in order to determine whether the two states will hybridize.
The effect of the kinetic term is
\begin{align}
-J (b_1^\dagger b_2 + b_2^\dagger b_1) |m\rangle &= -J \sqrt{(m+1)(N-m)} |m+1\rangle \nonumber \\
&\quad -J \sqrt{m(N-m+1)} |m-1\rangle.
\end{align}
The state $|m\rangle$ hybridizes with the state $|m+1\rangle$ when the first term above becomes comparable to the level spacing, i.e.
\begin{align}
& J \sqrt{(m+1)(N-m)} \sim (N-2m-1) U/2 \nonumber \\
& \Rightarrow u = \frac{NU}{J} \sim \frac{2N \sqrt{(m+1)(N-m)}}{N-2m-1}. \label{QuantumOnsetEqn}
\end{align}
The most excited state ($m=0$) has the highest level spacing and the effect of the kinetic term is the least for this state, hence they hybridize at the end.
Thus as $J$ is increased from zero, the energy levels begin to hybridize starting from the ground state.
Higher values of $J$ results in hybridization of more energetic eigenstates and the process continues until all the states are hybridized.
We confirm this picture and the above formula by a simple numerical test.
For the classical states the particle number fluctuations $\mathcal{F}$ is almost zero, as these states are eigenstates of the number operator.
We calculate this quantity as a function of $u$ and $\epsilon$ (Fig.~\ref{TwoSiteObservables}~(d)).
Quite remarkably the analytical result Eqn.~(\ref{QuantumOnsetEqn}) agrees almost perfectly with the contour at $\mathcal{F} = 1$ (not shown).
Thus the perturbation theory in this limit captures the onset of quantum effects quite well.
But this onset is not the localization transition itself, in that there is no sudden change in properties across this line.
The analysis presented here describes just the onset of quantum effects such that the particle number fluctuations becomes of order $1$, whereas the transition is accompanied by a very sharp change in this quantity.
A curious thing about this onset behavior is that it is not captured by other physical quantities we considered, whereas the transition itself is.

\bibliography{bibliography}
\end{document}